\documentclass[aps,prb,two column, superscriptaddress,amsmath,amssymb]{revtex4-2}
\usepackage{graphicx}
\usepackage{epsfig}
\usepackage{epstopdf}
\usepackage{amsmath}
\usepackage{amssymb}
\usepackage{amsthm}
\usepackage{amsfonts}
\usepackage{braket}
\usepackage{xcolor}
\usepackage[normalem]{ulem}
\usepackage{mhchem}
\usepackage{textcomp,mathcomp}
\usepackage{accents}
\graphicspath{{Images/}}

\begin{document}



\title{Magnetically coupled charge-transport crossover and giant negative magnetoresistance in iodine-incorporated Cr$_2$Se$_3$}
\setcounter{footnote}{1}

\author{Bikash Das}
\thanks{These authors contributed equally to this work.}
\email{bikashcgr1@gmail.com}
\affiliation{School of Physical Sciences, Indian Association for the Cultivation of Science, 2A \& B
Raja S. C. Mullick Road, Jadavpur, Kolkata - 700032, India}

\author{Shibnath Mandal}
\thanks{These authors contributed equally to this work.}
\affiliation{School of Physical Sciences, Indian Association for the Cultivation of Science, 2A \& B
Raja S. C. Mullick Road, Jadavpur, Kolkata - 700032, India}

\author{Kapildeb Dolui}
\affiliation{Lomare Technologies Limited, 6 London Street, London EC3R 7LP, United Kingdom}
\affiliation{Department of Physics, Indian Institute of Technology Tirupati, Tirupati, Andhra Pradesh 517619, India}


\author{Oleksi Laguta}
\affiliation{Central European Institute of Technology, Brno University of Technology, Purky\v nova 656/123, 61200
Brno, Czech Republic}

\author{Sambit Choudhury}
\affiliation{UGC-DAE Consortium for Scientific Research, Khandwa Road, Indore 452001, Madhya Pradesh, India}


\author{Suvadip Masanta}
\affiliation{Department of Physics, Bose Institute,Kolkata 700009, India}

\author{Tanima Kundu}
\affiliation{School of Physical Sciences, Indian Association for the Cultivation of Science, 2A \& B
Raja S. C. Mullick Road, Jadavpur, Kolkata - 700032, India}

\author{Jorge Andres Navarro Giraldo}
\affiliation{Central European Institute of Technology, Brno University of Technology, Purky\v nova 656/123, 61200
Brno, Czech Republic}

\author{Alexey Barinov}
\affiliation{Sincrotrone Trieste s.c.p.a., 34149 Basovizza, Trieste, Italy}

\author{Shibabrata Nandi}
\affiliation{Forschungszentrum Jülich GmbH, Jülich Centre for Neutron Science (JCNS-2) and Peter Grünberg Institut (PGI-4),
JARA-FIT, 52425 Jülich, Germany}

\author{Kai Rossnagel}
\affiliation{Ruprecht Haensel Laboratory, Deutsches Elektronen-Synchrotron DESY, 22607 Hamburg, Germany}
\affiliation{Institut für Experimentelle und Angewandte Physik, Christian-Albrechts-Universität zu Kiel, 24098 Kiel, Germany}

\author{Sanjoy Kr Mahatha}
\affiliation{UGC-DAE Consortium for Scientific Research, Khandwa Road, Indore 452001, Madhya Pradesh, India}

\author{Rajib Mondal}
\affiliation{UGC-DAE Consortium for Scientiﬁc Research, Kolkata Centre, Bidhannagar, Kolkata 700 106, India}

\author{Petr Neugebauer}
\affiliation{Central European Institute of Technology, Brno University of Technology, Purky\v nova 656/123, 61200
Brno, Czech Republic}

\author{Subhadeep Datta}
\email{sspsdd@iacs.res.in}

\affiliation{School of Physical Sciences, Indian Association for the Cultivation of Science, 2A \& B
Raja S. C. Mullick Road, Jadavpur, Kolkata - 700032, India}



\begin{abstract}
{We report the synthesis and comprehensive investigation of iodine-incorporated $Cr_2Se_3$, a non-van der Waals quasi-two-dimensional magnetic material, using structural, magnetic, transport, spectroscopic, and first-principles methods. Magnetization and electron spin resonance measurements reveal an antiferromagnetically ordered state below $T_N \approx 52$ K. At higher temperatures, a second, broader anomaly emerges near $T^* \approx 150$ K, coinciding with a shallow minimum in the temperature-dependent resistivity that resembles a metal-to-insulator-like crossover. Hall measurements indicate predominantly hole-like conduction at high temperatures ($\geq$ 150 K), while the nonlinear Hall response below T*, together with an anomaly in the third-harmonic electrical signal, suggests the emergence of mobility-dependent multichannel and spatially inhomogeneous transport. A large, non-saturating negative magnetoresistance reaching approximately -78\% at 25 K and 12 T further demonstrates strong coupling between charge transport and the magnetic state. Temperature-dependent Raman spectroscopy reveals no symmetry-changing structural transition near T*, whereas angle resolved photoemission spectroscopy (ARPES) measurements show no major reconstruction of the electronic structure occupied across the crossover. Taken together, these results identify the high-temperature anomaly as a broad magnetically coupled transport crossover arising from the interplay of short-range magnetic correlations, chemical disorder, and redistribution among competing conduction channels. These findings demonstrate that anion incorporation provides an effective route to tuning the coupled electronic and magnetic properties of non-layered Cr$_2$Se$_3$, establishing this system as a promising platform for investigating correlated transport and emergent spintronic functionalities in transition-metal chalcogenides.}

\end{abstract}

\maketitle

\section{Introduction}

The precise manipulation of magnetic properties in quasi-two-dimensional (quasi-2D) materials has emerged as a central challenge for controlling their ground states and realizing novel quantum functionalities. In these systems, the delicate interplay among charge, spin, and lattice degrees of freedom enables a variety of emergent phenomena, including carrier-induced magnetic ordering, electronic phase transitions, and unconventional magnetotransport responses \cite{guo2014large,chen2022antiferromagnetic}. Introducing charge carriers through chemical doping can substantially modify the exchange interactions and magnetic anisotropy, driving the formation of inhomogeneous magnetic states such as correlated or ferromagnetic regions embedded within a paramagnetic or antiferromagnetic background. Consequently, understanding how carrier doping modifies magnetic correlations and stabilizes emergent magnetic states remains an important challenge for both fundamental condensed matter physics and future spintronic applications. Anion doping is advantageous due to the lower energy barrier for incorporation, as negatively charged anions, located on the material’s surface, reduce the activation energy required for substitution compared to cation doping \cite{PhysRevLett.134.066401}. 
Additionally, anion doping avoids intercalation issues prevalent in cation doping, where cations accumulate in interlayer spaces due to electrostatic attraction to surrounding anions, increasing lattice disorder and raising the energy cost of dopant incorporation \cite{guo2014large,zheng2020boosting,hu2008evidence}. 
Consequently, anion doping offers more direct approach to modulate electronic and magnetic properties, enabling efficient exploration of fundamental material characteristics without the stringent requirements of advanced thin-film processing \cite{schlom2008thinfilm}. Moreover, unlike cation substitution, anion substitution minimally perturbs lattice symmetry while selectively modulating orbital overlap and spin exchange pathways.




In many doped chalcogenide systems, substituting one chalcogen or halogen modifies the transition-metal cation environment by altering bond angles and bond lengths, thereby introducing small structural distortions that modify the magnetic exchange interactions in accordance with the Goodenough - Kanamori rules \cite{goodenough1955theory,kanamori1959superexchange}. In recent anion-doped quasi-2D magnetic systems, Li \textit{et al.} \cite{li2025modulation} demonstrated that substituting S/Se at the Te site of Cr$_{\mathrm{2}}$Te${_\mathrm{3}}$ enhances the magnetic interaction strength due to the smaller ionic radii of the dopant anions. Basnet \textit{et al.} \cite{basnet2022controlling} examined the tunability of the antiferromagnetic transition temperature in MnPS$_{\mathrm{3-x}}$Se${_\mathrm{x}}$ and NiPS$_{\mathrm{3-x}}$Se${_\mathrm{x}}$ (0 $\leq$ x $\leq$ 3) through controlled modification of their magnetic anisotropy. They observed opposite doping trends in the two systems, likely originating from their distinct dominant exchange pathways: direct Mn–Mn exchange in MnPS$_{\mathrm{3}}$ versus Ni–S–Ni superexchange in NiPS$_{\mathrm{3}}$. Abramchuk et al. \cite{abramchuk2018controlling} further showed that Br substitution in CrCl${_\mathrm{3}}$ drives a transition from antiferromagnetic to ferromagnetic order while simultaneously modifying the magnetic easy axis. In addition, Tartaglia et al. \cite{tartaglia2020accessing} reported the emergence of a frustrated magnetic interactions in CrCl$_{\mathrm{3-x-y}}$Br${_\mathrm{x}}$I${_\mathrm{y}}$, arising from the tuning of ligand spin–orbit coupling. Doping a divalent chalcogenide with a monovalent halogen together with intrinsic vacancies or non-stoichiometry during growth further modifies the local crystal field and chemical bonding, altering orbital hybridization and electronic bandwidth. Such changes can favor carrier localization, leading to Mott-like or polaronic electronic behavior.

Cr-based Cr$_{\mathrm{2}}$X${_\mathrm{3}}$ (X = S, Se, or Te) systems, a family of non-van der Waals quasi-2D and air-stable magnetic materials, provide a promising platform for investigating tunable magnetic properties \cite{lu2024out,roy2015perpendicular}. These materials feature a unique structure with five atomic layers in a X--Cr--X--Cr--X configuration, with Cr atoms at the center of [CrSe$_{6}]^{9-}$ octahedra. The magnetic properties of Cr$_{\mathrm{2}}$Se${_\mathrm{3}}$ remain debated, with stoichiometric Cr$_{\mathrm{2}}$X${_\mathrm{3}}$ exhibiting non-collinear antiferromagnetism (AFM) and weak ferromagnetism (FM) below the Néel temperature ($T_N$) \cite{gebredingle2022first,wu2020magnetotransport,adachi2007electrical}. First-principles calculations further predict that monolayer Cr$_{\mathrm{2}}$Se${_\mathrm{3}}$ is a ferromagnetic semiconductor, highlighting the strong sensitivity of its magnetic ground state to reduced dimensionality and electronic structure \cite{he2021two}. The magnetic transition in Cr$_{\mathrm{2}}$Se${_\mathrm{3}}$ is coupled to a first-order structural phase transition, as evidenced by temperature-dependent thermal expansion measurements, which may influence electronic properties through lattice-mediated effects \cite{ohta1994thermal}. In thermoelectric applications, polycrystalline Cr$_{\mathrm{2}}$Se${_\mathrm{3}}$ achieves a figure of merit $ZT \approx 0.25$ at approximately 600 K, highlighting its multifunctional potential \cite{zhang2018structure}. Notably, Te-substituted Cr$_{\mathrm{2}}$Se${_\mathrm{3}}$ manifests as an antiferromagnetic semiconductor, contrasting with the metallic nature of pure Cr$_{\mathrm{2}}$X${_\mathrm{3}}$, suggesting doping-induced modulation of electronic states .

In this report, we successfully synthesized single crystals of iodine-incorporated Cr$_{\mathrm{2}}$Se${_\mathrm{3}}$ using the chemical vapour transport (CVT) technique. We primarily investigated the structure and phase purity of the samples through X-ray diffraction (XRD) and high-resolution transmission electron microscopy (TEM). Successful iodine incorporation was confirmed by energy-dispersive X-ray spectroscopy (EDS) and X-ray photoelectron spectroscopy (XPS). Both DC and AC magnetization measurements, ESR spectroscopy along with theoretical studies, indicate that the sample exhibits a predominantly antiferromagnetic (AFM) order below 52 K. A second, broader anomaly appears near T* $\approx$ 150 K and is accompanied by a shallow minimum in the temperature-dependent resistivity. This high-temperature anomaly corresponds to a sign change of the temperature coefficient of resistivity. ARPES measurements reveal a predominantly semiconducting-like occupied electronic structure without a major reconstruction across the transport crossover. In particular, at 25 K, the crystal exhibits a giant, non-saturating negative magnetoresistance of -78\%, which surpasses the magnitude observed in the initially reported giant magnetoresistance (GMR) material, the Fe/Cr/Fe heterostructure \cite{gijs1992magnetoresistance}.
\section{Methods}

Single crystals of iodine-incorporated Cr$_{\mathrm{2}}$Se$_{\mathrm{3}}$ were grown using standard CVT technique. Elemental Cr (99.999\%), Se (99.999\%) and I (99.99\%) were mixed and sealed under a vacuum of $~10^{-5}$ mbar. Then the sealed tube was placed in a two-zone tube furnace, with the hot end set at 850 °C and the cold end at 700 °C. The temperature gradient was maintained for 5-7 days. After the reaction, the tube was cracked and the millimeter-sized crystals were collected from the hot and middle regions of the tube. 
Powder X-ray diffraction (PXRD) data were taken using a
Rigaku SmartLab (Cu $K_\alpha$ radiation, $\lambda = 1.5406$ Å) diffractometer. TEM images were captured on a JEOL-JEM-F200 electron microscope with an accelerating voltage of 200 kV. 

Temperature dependent dc/ac susceptibility and specific heat measurements were conducted using a physical property measurement system (PPMS) (Quantum Design Inc. DynaCool 9T). Temperature dependent dc/ac resistivity measurement were performed using a ARS cryostat. Magneto transport measurements were carried out utilizing a 15 tesla cryogen free measurement system (Cryogenic Limited, U.K.). 


Microwave absorption measurements were performed using a home-built broadband electron paramagnetic resonance (EPR) spectrometer operating at 90--1100\,GHz microwave frequencies, magnetic fields of -16--16\,T and temperatures in the range of 2--320\,K \cite{Laguta2021,Sojka2022,Sedivy2023}. The sample was mounted on the reflecting mirror of a non-resonant holder so that the external magnetic field was normal to the sample plane (H//c). For better sensitivity, phase-sensitive detection with modulation of the magnetic field (10\,Oe at 3\,kHz) was used, which also resulted in the derivative of the absorption spectrum $\frac{d\chi}{dH}$.

To investigate the vibrational properties, Raman spectroscopy measurements were performed using a micro-Raman system equipped with a spectrometer (LabRAM HR, Jobin Yvon) and a Peltier-cooled CCD detector. A He–Ne laser with a wavelength of 633 nm and a spot size of approximately 1 $\mu$m was used as the excitation source.

Electronic structure calculations for the bulk Cr$_2$Se$_3$ were carried out within the framework of density functional theory (DFT) using the screened hybrid functional of Heyd-Scuseria-Ernzerhof (HSE06)~\cite{heyd2003hybrid} as implemented in QuantumATK package~\cite{smidstrup2019quantumatk}. Norm-conserving fully relativistic pseudopotentials of the type PseudoDojo~\cite{smidstrup2019quantumatk, van2018pseudodojo} for describing electron-core interactions; and the Pseudojojo (medium) numerical linear combination of atomic orbitals (LCAO) basis set~\cite{van2018pseudodojo}. The energy mesh cutoff for the real-space grid was chosen as 101 Hartree, and the k-point grid 6$\times$6$\times$2 was used for the self-consistent calculations. Density of states were obtained by integrating over 12$\times$12$\times$4 $k$-point mesh. Spin-orbit coupling was not included in the electronic structure calculations.

Soft X-ray ARPES measurements were performed at the ASPHERE-III end-station of the P04 beamline at PETRA III, DESY (Germany), using a focused photon beam with a spot size of $\sim$ 10 $\mu$m. The spectra were recorded at 87 K and 250 K in normal emission geometry using a Scienta-Omicron DA30 electron energy analyzer. The total energy resolution was approximately 80 meV with a momentum resolution better than 0.02 $Å^{-1}$.

\section{Results}

\subsection{Structure}

\begin{figure*}[h!]
   \centering
    \includegraphics[width=0.9\linewidth]{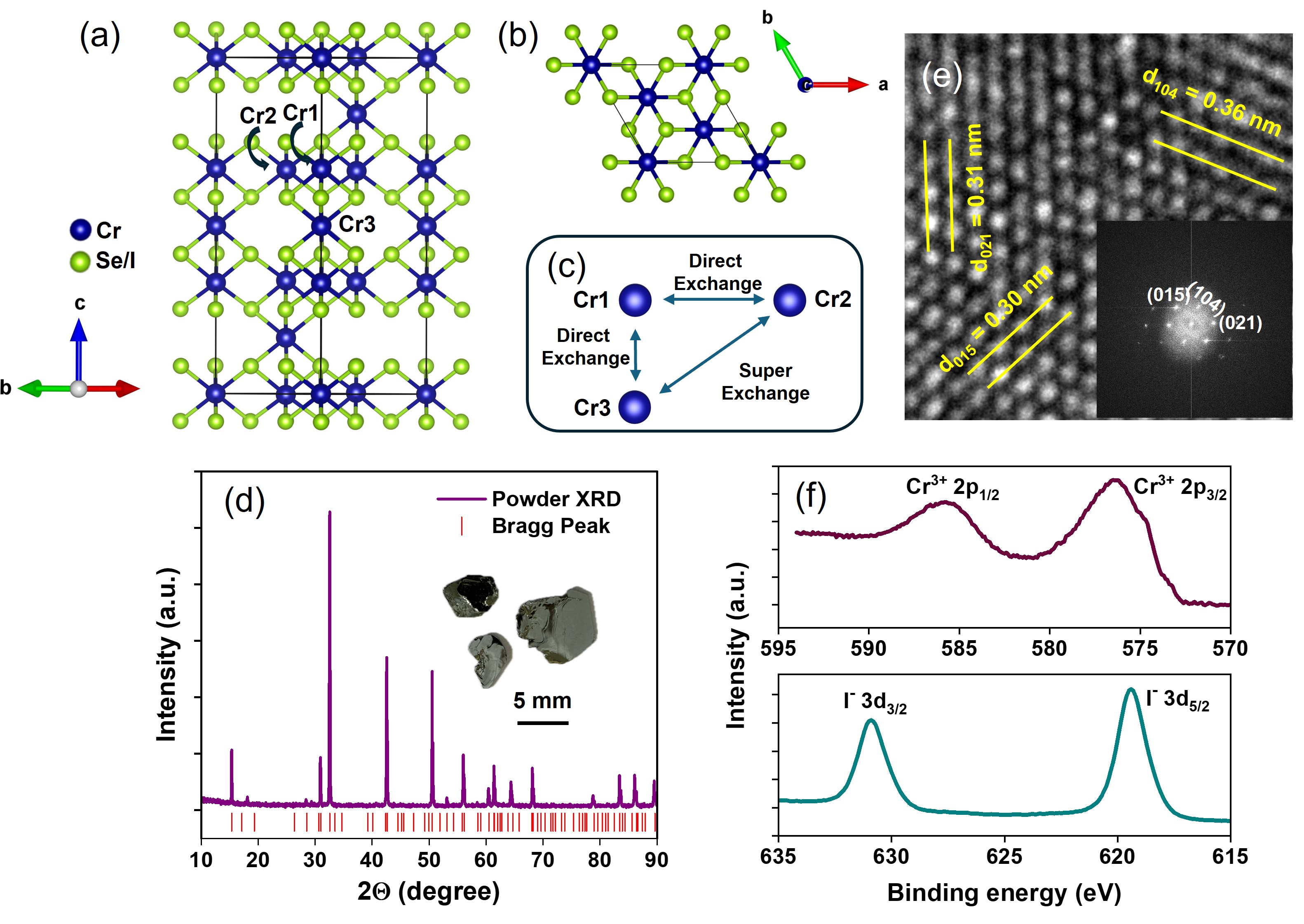}
    \caption{(a), (b) Crystal structure of Cr$_{\mathrm{2}}$Se$_{\mathrm{3}}$. (c) Different direct and superexchange interactions between the Cr atoms in the unit cell. (d) Powder XRD pattern of I@Cr$_{\mathrm{2}}$Se$_{\mathrm{3}}$ at room temperature. The Bragg positions corresponding to pure Cr$_{\mathrm{2}}$Se$_{\mathrm{3}}$ are also indicated. Inset: millimeter-sized CVT-grown I@Cr$_{\mathrm{2}}$Se$_{\mathrm{3}}$ crystals. (e) High-resolution TEM image of doped crystal. Three different lattice planes, (021), (015) and (104) with their corresponding d-spacings 0.31, 0.30 and 0.36 nm, respectively are also indicated in the figure. Inset: SAED of the crystal structure. The specified planes are assigned with the respective first-order spots in the SAED pattern. (f) Core-level XPS data of Cr 2p and I 3d.}
    \label{structure}
\end{figure*}

Cr$_{\mathrm{2}}$Se$_{\mathrm{3}}$ belongs to the family of metal-deficient NiAs-type structure and crystallises in rhombohedral \textit{R}-3 space group with lattice parameters $a=6.33$ Å, and $c=17.66$ Å. \cite{burn2019cr2te3}. 
The crystal structure is formed by alternating CrSe and Cr-deficient $Cr_{1/3}Se$ layers stacked vertically. The crystal structure contains three distinct Cr sites, labelled Cr1, Cr2 and Cr3, as shown in Figure~\ref{structure}(a). These Cr atoms are surrounded by an octahedral environment formed by Se atoms. The varying Cr sites correspond to three types of irregular $[CrSe6]^{9-}$ octahedra, characterized by Cr-Se-Cr bond angles that deviate from 90°. The octahedra are edge-haring in the ab plane and face-sharing along the out-of-plane direction. Within the unit cell, Cr3 atoms lack direct neighbors in the ab plane but are flanked by Cr1 atoms in the neighboring up and down layers. Cr2 atoms neighbor with Cr1 within the same layers but lack adjacent Cr atoms along the c-axis. Layers with Cr2 and Cr1 are fully occupied, while layers with Cr3 have vacancies \cite{gebredingle2022first,zhong2023stoner}. 
The magnetic ground state of Cr$_{\mathrm{2}}$Se$_{\mathrm{3}}$ arises from the interplay between direct-exchange and superexchange mediated by Se atoms. The interlayer and intralayer magnetic exchanges within Cr$_{\mathrm{2}}$Se$_{\mathrm{3}}$ are illustrated in Figure~\ref{structure}(c). Theoretical calculations indicate that Cr$_{\mathrm{2}}$Se$_{\mathrm{3}}$ exhibits a fully compensated ferrimagnetic ground state, which can be transitioned to ferromagnetic by substituting Te or Se \cite{gebredingle2022first,burn2019cr2te3}. Consequently, it remains an open question what ground state may be achieved through any anion doping into the system.

Figure~\ref{structure}(a) and (b) illustrate the structure of Cr$_{\mathrm{2}}$Se$_{\mathrm{3}}$ from two mutually perpendicular directions, with the different Cr sites labeled. After growing the single crystals, a few of them were ground into powder for performing PXRD, as shown in Figure \ref{structure}(d). The observed reflections can be indexed using the rhombohedral Cr$_{\mathrm{2}}$Se$_{\mathrm{3}}$ structure, and no additional crystalline phase is detected within the resolution of the laboratory PXRD measurement. A representative image of the as-grown crystals is provided in the inset of \ref{structure}(d). The qualitative elemental composition was determined by EDS analysis. The observed atomic ratio was $Cr:Se:I=40:55:5=2:2.75:0.25$, indicating 8.33\% iodine incorporation. Structural characterization of I@Cr$_{\mathrm{2}}$Se$_{\mathrm{3}}$ was also performed via TEM. Figure \ref{structure}(e) depicts a high-resolution TEM image of the crystal. 
The measured lattice spacings are $0.30 \pm 0.01$, $0.31 \pm 0.01$, and $0.36 \pm 0.01$ nm corresponding to the (015), (021), and (104) planes, respectively. The estimated lattice spacing are slightly higher than the theoretical values which may be associated with iodine incorporation; however, the present measurements do not establish the specific crystallographic site occupied by iodine. A selected area electron diffraction (SAED) pattern is presented in the inset of Figure~\ref{structure}(e). First-order bright spots in the fast Fourier transform (FFT) pattern correlate with various Bragg planes (marked in the SAED image), consistent with the d-spacing and angles as observed in TEM study. Figure \ref{structure}(f) shows the XPS spectra for the Cr 2p and I 3d regions
of I@Cr$_{\mathrm{2}}$Se$_{\mathrm{3}}$ single crystal. The peaks at 576.4 eV and 585.8 eV correspond to the doublet peaks of Cr 2p \cite{gupta2013static}, while the peaks at 619.4 eV and 630.9 eV are attributed to I 3d \cite{guo2014large}. The XPS results confirm the presence of iodine in the sample.

\subsection{Magnetization}

\begin{figure*}[h!]
   \centering
   \includegraphics[width=1.0\linewidth]{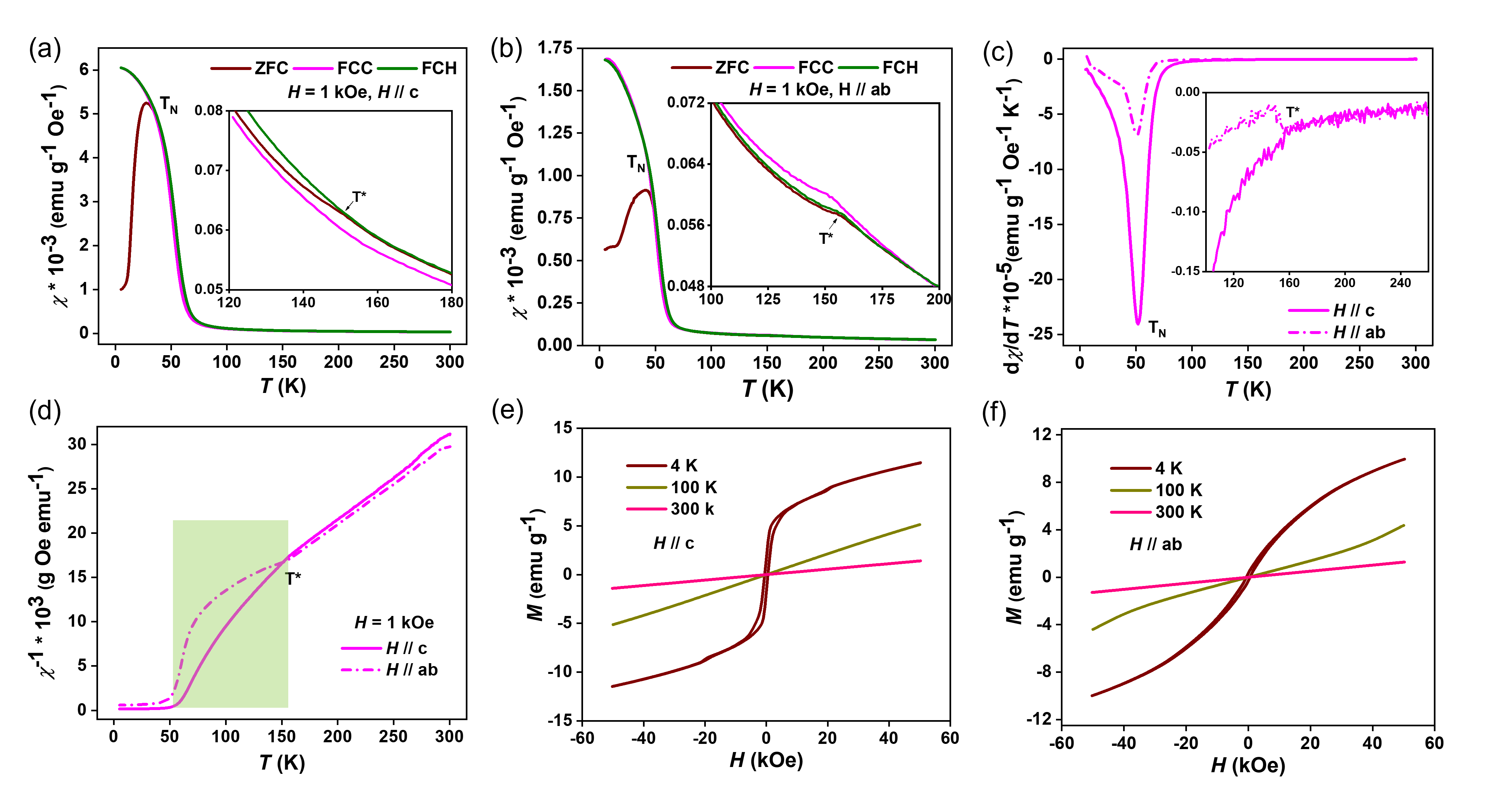}
    \caption{Temperature dependent magnetic susceptibility $\chi$ measured in standard ZFC-FCC-FCH protocol at an applied magnetic field 1 kOe for (a) H//c and (b) H//ab direction of the crystal. 
Inset: zoomed-in view of the high temperature transition observed at (a) 150 K for H//c and (b) 156 K for H//ab, respectively. 
(c) Derivatives of the FCC plot $\frac{d\chi}{dT}$. The continuous line corresponds to H//c, while the dashed line represents H//ab. The dip of the plots indicates the $T_N=52$ K. Inset shows the high temperature regions of the two derivatives where the transition, T* is shown. (d) $\chi^{-1}-T$ plot for both directions. The shaded region indicates the non analytical magnetization behaviour below T*. M-H data at three different temperatures for (e) H//c and (f) H//ab.  
}
    \label{DC_chi}
\end{figure*}

Temperature dependent dc magnetization data were collected using conventional zero field cooling (ZFC), field cooled cooling (FCC), and field cooled heating (FCH) protocols in both out-of-plane (H//c) and in-plane (H//ab) directions of the crystal. From the $\chi(T)$ data as shown in Figure~\ref{DC_chi}(a) and (b), a magnetic transition is observed at an applied magnetic field of 1 kOe in both ZFC and FC curves. A significant bifurcation between ZFC and FCC suceptibility in both directions of the crystal is noted. This large bifurcation can be attributed to magnetic anisotropy, formation of FM clusters, spin frustration, or a possible spin glass state in the sample \citep{dho2002reentrant,nagpal2022evidence}. In both ZFC data, the magnetic susceptibility display a notable peak (31 K for H//c and 40 K for H//ab). However, the FCC/FCH susceptibilities indicate a steady increase with decreasing temperature, as illustrated in Figure~\ref{DC_chi}(a) and (b). Also, a higher $\chi$ value for H//c compared to H//ab confirms the sample exhibits out-of-plane magnetic anisotropy. The derivatives of the FCC susceptibility ($\frac{d\chi}{dT}$) are plotted in Figure~\ref{DC_chi}(c). Prominent dips at 52 K in both $\frac{d\chi}{dT}$ data imply the long-range magnetic transition temperature of the sample. 
The observed transition temperature is in close agreement with those reported for pristine Cr$_{\mathrm{2}}$Se$_{\mathrm{3}}$ and Ti@Cr$_{\mathrm{2}}$Se${_\mathrm{3}}$ samples \cite{wu2020magnetotransport,zhang2021colossal}. However, an additional anomaly is observed at a comparatively higher temperature ($T^*\approx 150$ K), as indicated in the zoomed-in $\chi$ vs. T data shown in the inset of Figure~\ref{DC_chi}(a) and (b). A kink-like transition is found at 150 K (and 156 K) for H//c (and H//ab) direction. This anomaly is also reflected in the zoomed-in $\frac{d\chi}{dT}$ plot (see inset of Figure~\ref{DC_chi}(c)). We also plotted the inverse susceptibility ($\chi^{-1}$) as a function of temperature, presented in Figure~\ref{DC_chi}(d). In the temperature region $200\leq T\leq 300$ K, $\chi^{-1}$ vs. T data follows the modified Curie-Weiss (C-W) law as follows \cite{mugiraneza2022tutorial}:
\begin{equation}
\begin{aligned}
\chi(T)= \chi_0 + \frac{C}{T- \theta_p}
\end{aligned}
\label{eq:hall}
\end{equation}
, where $\chi_0$ is the temperature-independent susceptibility, C is the Curie constant and $\theta_p$ is the C-W temperature. The extracted negative values (-92 $\pm$ 7 K for H//c and -66.4 $\pm$ 7 K for H//ab) of $\theta_p$ indicate prominent AFM interactions between the neighbouring magnetic atoms in the compound. 
An anomalous deviation from the linear Curie--Weiss behavior is observed above the long-range ordering temperature, extending up to a characteristic temperature $T^*$ for both field orientations, as shown in Figure~\ref{DC_chi}(d). Such non-analytic behavior is commonly associated with the formation of ferromagnetic clusters within the paramagnetic matrix prior to the establishment of long-range magnetic order and may also indicate a Griffiths-like regime in correlated materials \cite{griffiths1969nonanalytic,saha2023short}. To examine this possibility, we analyzed the inverse susceptibility using the Griffiths-phase scaling relation. However, no physically reasonable and consistent fit was obtained over the relevant temperature range. Notably, the $T^*$ anomaly is strongly suppressed under high magnetic fields ($H > 10$ kOe) for H//c, whereas it remains visible for H//ab (see Figure S1(a) and (b)). Since the c-axis is the magnetic easy axis, a field applied along H//c more effectively polarizes the correlated moments and suppresses their contribution near T*. For H//ab, the weaker polarization allows the anomaly to remain visible at higher fields. 

Figure \ref{DC_chi}(e)-(f) display the magnetic field-dependent magnetization for both H//c and H//ab directions of the crystal. At 4 K, I@Cr$_{\mathrm{2}}$Se$_{\mathrm{3}}$ shows weak magnetic hysteresis loops at low fields and remains unsaturated at high fields in both directions. The hysteresis suggests the presence of a small irreversible magnetic component within the magnetic state, possibly arising from spin canting or uncompensated moments associated with vacancies or inequivalent magnetic sites. 
 Notably, a field-induced transition occurs around 20 kOe only for H//c, suggesting a possible field-induced metamagnetic transition \cite{das2023metamagnetism}. At 100 K, the M-H curve shows linear behavior for H//c, while the non-linear M-H curve for H//ab indicates the presence of anisotropic short-range magnetic correlations above $T_N$. At 300 K, both M-H datasets exhibit linear paramagnetic behavior, as shown in Figure \ref{DC_chi}(e) and (f). 

\begin{figure*}[h!]
   \centering
    \includegraphics[width=1\linewidth]{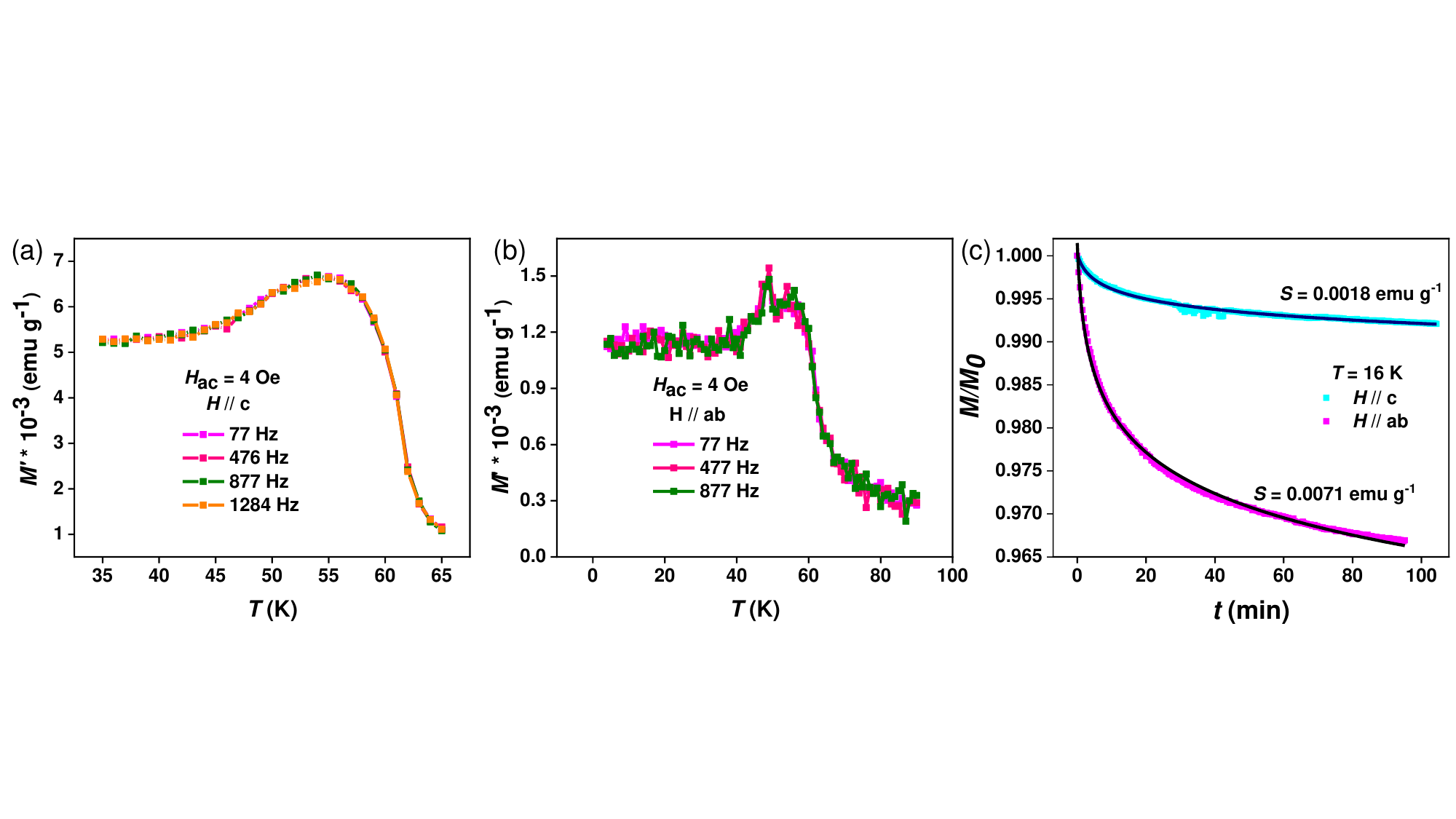}
    \caption{(a) Temperature-dependent real part of ac magnetization measured at different frequencies for (a) H//c and (b) H//ab. (c) Isothermal remanent magnetization (IRM) measured in two different directions of the applied magnetic field relative to the sample. }
    \label{AC_chi}
\end{figure*}

To test for spin-glass like disorder in I@Cr$_{\mathrm{2}}$Se$_{\mathrm{3}}$, we conducted AC susceptibility measurements on the single crystals. Figure \ref{AC_chi}(a) and (b) represent the frequency-dependent real part of ac magnetic moment of I@Cr$_{\mathrm{2}}$Se$_{\mathrm{3}}$ as a function of temperature when an oscillating magnetic field (4 Oe) was applied along two mutually perpendicular directions of the crystal. The absence of a change in the long-range transition temperature ($T_N$) with varying frequencies indicates that the sample does not exhibit characteristics of a spin glass system \cite{das2023emergence}. 
Isothermal remanent magnetizations ($M_{IRM}$) of the sample are also plotted as a function of time in both H//c and H//ab configurations at 16 K, as shown in Figure \ref{AC_chi}(c). The 
$M_{IRM}$ as a function of time (t) can be fitted as \cite{majumdar1999magnetic}: 
\begin{equation}
\begin{aligned}
{M}_{{\rm{IRM}}}\left( t \right)\ = {M}_0 - S\left( T \right){\rm{ln}}\left( {1 + \frac{t}{{{t}_0}}} \right)
\end{aligned}
\label{eq:hall}
\end{equation}, where $M_0$ and S(T) are the initial magnetization at t = 0 and magnetic viscosity, respectively. The coefficient $t_0$ depends on
the measuring conditions of the magnetometer and has limited physical significance. A slow relaxation of $M_{IRM}$ over time was observed for both field orientations, where $M_{IRM}$ decreases by 0.9\% and 3.4\% of its initial value after 100 minutes for H//c and H//ab directions, respectively. The calculated value of $S(T)\approx 0.0018$ emu $g^{-1}$ (0.0071 emu $g^{-1}$) for H//c (H//ab) from the fitting is comparable to the previously reported value of S(T) for Y-doped $Gd_2PdSi_3$ alloys \cite{majumdar1999magnetic}, which are free from the spin-glass state at low temperatures.

Figure S2(a) shows the ESR spectra (measured at the microwave frequency of 95\,GHz)  below and around $T_N$. Below the temperature $T_N$, the spectra consist of two features. A relatively strong and broad peak with temperature-dependent position and intensity is a typical signature of antiferromagnetic resonance (AFMR) \cite{Foner1963}. It is observable only below $T_N$ supporting the results of magnetization measurements. In the temperature range $T_N<T<T^*$, the ESR spectra resemble those of ferromagnetic resonance (FMR) (Figure S2(b)), along with a distinct change in the resonance response near T*, as described in detail in the supporting information file. 

\subsection{Magnetotransport}

\begin{figure*}[h!]
   \centering
    \includegraphics[width=1\linewidth]{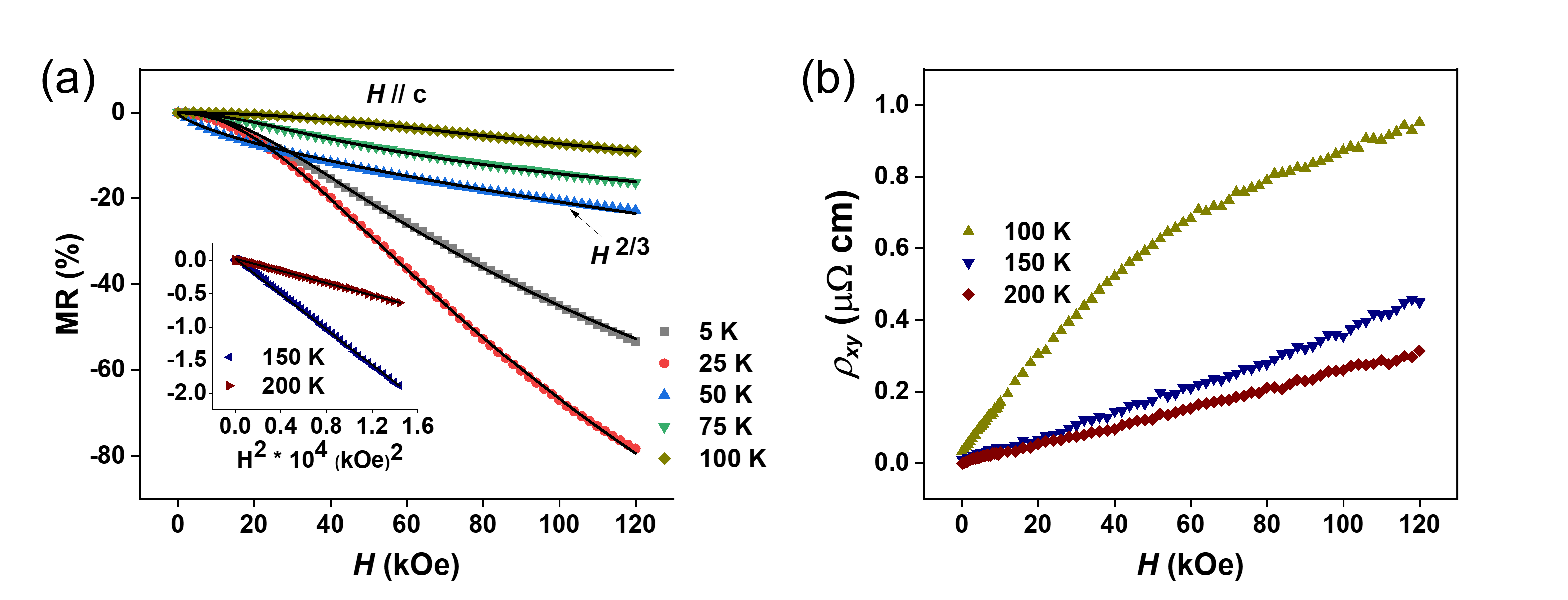}
    \caption{(a) MR measurement: giant negative magnetoresistance at various temperatures for H//c, along with appropriate fittings using suitable models for different temperature ranges (below $T_N$, near $T_N$, and above $T_N$). Above 100 K, the data are well described by the model adapted from Ref. \cite{khosla1970magnetoresistance}, except at 50 K, where a $H^\frac{2}{3}$ dependence is observed. 
    The insets display the high-temperature magnetoresistance behavior where $MR\propto -H^2$. (b) Hall measurement: at 150 K and 200 K linear $\rho_{xy}$ vs. H dependence suggests the dominating charge carriers are hole, whereas, at 100 K, the sublinear $\rho_{xy}$ vs. H implies the existence of more than one type of charge carrier following the classical two-band model. }
    \label{MR}
\end{figure*}

Magnetic field dependent resistance (MR) at various temperatures (for H//c) is presented in Figure~\ref{MR}(a). During the measurements, transverse magnetoresistance (TMR) was recorded, with the current applied perpendicular to the magnetic field. The percentage of MR was calculated using the relation: 

\begin{equation}
\begin{aligned}
MR(\%)= \frac{\rho_{xx}(H)-\rho_{xx}(0)}{\rho_{xx}(0)}*100
\end{aligned}
\label{eq:hall}
\end{equation}

We observed a giant non-saturating negative magnetoresistance of $-78\%$ 
at 25 K with an applied magnetic field 120 kOe. At 5 K, this negative MR value reduces to $-53\%$ 
due to the higher resistivity of the crystal compared to 25 K (low temperature resistivity behavior is discussed in the next section). Above 25 K, the MR signal decreases 
as the temperature increases, reaching -0.6\% at 200 K, as shown in the inset of Figure~\ref{MR}(a).

Next, we attempted to fit our MR data at different temperatures with proper model. The well-known second-order terms of the s-d Hamiltonian, which lead to $MR\propto -H^2$ dependence \cite{yosida1957anomalous,toyozawa1962theory}, do not fully explain the observed MR data. We fitted our low-temperature data (5 - 100 K, excluding 50 K) using a model originally adapted by Khosla and Fischer for In-doped CdS semiconducting single crystals \cite{khosla1970magnetoresistance}:
\begin{equation}
\begin{aligned}
MR= -a^2 ln (1+b^2H^2) + \frac{c^2 H^2}{1+d^2H^2}
\end{aligned}
\label{eq:hall}
\end{equation}
The last term represents a well-known positive MR contribution derived from a two-band model. Parameters c and d are associated with the conductivities and mobilities of these two bands. Conversely, the parameters a and b involved in the negative MR contribution are defined as \cite{zapata2019sd}:
\begin{equation}
\begin{aligned}
a^2= a_1^2[S(S+1)+<M^2>]
\end{aligned}
\label{eq:hall}
\end{equation}
and
\begin{equation}
\begin{aligned}
b^2=[1+4S^2\pi^2\left(\frac{2J\rho_F}{g}\right)^4]\frac{g^2\mu_B^2}{\left(\alpha k_BT\right)^2}
\end{aligned}
\label{eq:hall}
\end{equation}

Here, g represents the Landé g-factor, $\mu_B$ denotes the Bohr magneton, S signifies the spin of the localized moment, J is the exchange integral arising from the interaction between the itinerant electron and the localized magnetic moment, $\rho_F$ indicates the density of states at the Fermi energy level, and $\alpha$ is a numerical factor of the order unity. 
We also observed a $H^\frac{2}{3}$ variation of MR near $T_N$, suggesting enhanced magnetic fluctuations following the theory predicted by Ref. \cite{yamada1973magnetoresistance} \textcolor{red}. Moreover, at 150 and 200 K, a -$H^2$ variation of the data is evident, following the s-d model \cite{yosida1957anomalous,toyozawa1962theory} (see insets of the Figure~\ref{MR}(a)). 
The negative MR observed significantly above the long range magnetic transition temperature can be linked to the possible ferromagnetic correlations, similar to those found in the van der Waals ferromagnet CrSiTe$_3$, as confirmed by neutron scattering experiment \cite{williams2015magnetic}. 
Other possible mechanisms such as weak localization \cite{qu2024observation}, chiral anomaly \cite{balduini2024intrinsic} or Kondo screening \cite{xue2018large} can be ruled out as the origin of the observed negative MR. Negative MR value for weak localization typically saturates at a lower magnetic field at low temperatures, which is not observed here. The absence of any detectable change in the MR with the current applied parallel to the magnetic field signifies that the effect is not linked to the chiral anomaly. Moreover, the observed very large negative MR is much stronger than that typically seen in Kondo systems.

Hall measurements were conducted at an applied magnetic field up to 120 kOe, where the transverse voltage $V_{xy}$ is recorded perpendicular to the current (I) direction. The Hall resistivity $\rho_{xy}$ is defined as $\rho_{xy} = (V_{xy}/I)t$, where t is the sample thickness. Hall resistivity $\rho_{xy}$ are plotted using the formula 
\begin{equation}
\begin{aligned}
\rho_{xy}=\frac{1}{2}[\rho_{xy}(+H)-\rho_{xy}(-H)]
\end{aligned}
\label{eq:hall}
\end{equation}
to eliminate any MR contribution in the data (see Figure~\ref{MR}(b)) \cite{bera2023anomalous}.  Positive Hall slope at high temperatures (150 K - 200 K) suggests the holes are the majority carriers in the system. At 200 K, the linear slope of the $\rho_{xy}$ vs. H curve indicates a hole concentration of $1.43*10^{22}$ $cm^{-3}$. At 100 K, $\rho_{xy}$ exhibits sublinear character, with a change in slope near 40 kOe, possibly indicating the presence of more than one type of charge carrier, consistent with the classical two-band model \cite{chatterjee2023unusual}. 

\subsection{Resistivity and heat capacity}

\begin{figure*}[h!]
   \centering
   \includegraphics[width=0.8\linewidth]{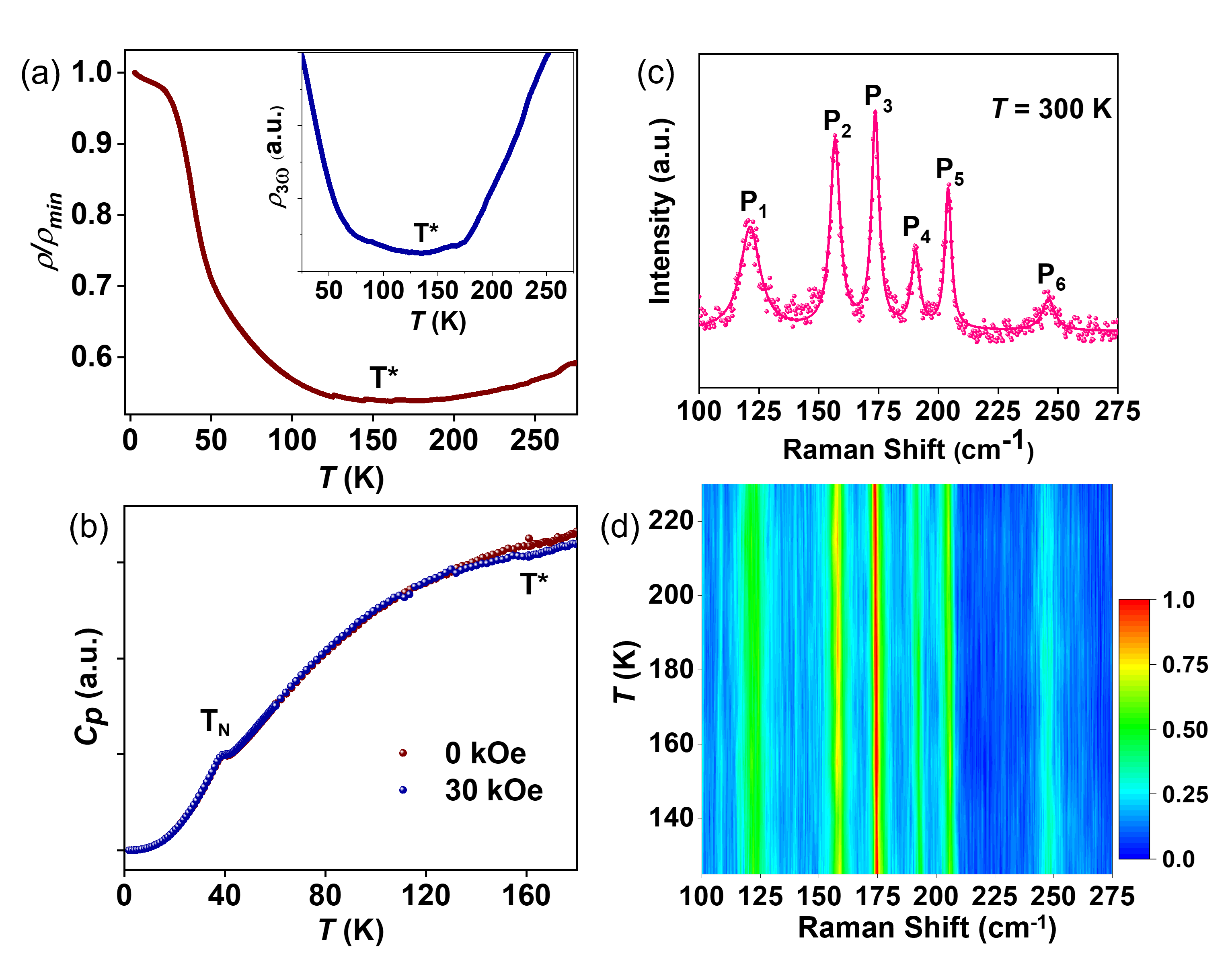}
      \caption{(a) Temperature dependent normalized resistivity $\left(\frac{\rho}{\rho_{min}}\right)$ data at zero magnetic field. Inset: third-harmonic resistivity as a function of temperature. Both datasets indicate the transition associated with the crossover near T*. (b) Temperature dependent specific heat ($C_p$) at two magnetic fields, 0 kOe and 30 kOe. Two transitions are indicated by $T_N$ and T*. (c) Raman spectrum at 300 K. (d) Contour plot of the Raman spectra collected in the temperature range 120-230 K. The color contrast indicates the normalized intensity scale.} 
    \label{BMP}
\end{figure*}

Figure \ref{BMP}(a) displays the temperature-dependent zero-field resistivity data, normalized to the resistivity at the lowest measured temperature ($\approx 3$ K). Initially, as the sample cools, it exhibits typical metallic behavior. However, below 162 K, it starts to increase with decreasing temperature, producing a broad minimum. Although this behavior resembles a metal-to-insulator-like transition \cite{adler1967semiconductor,banik2024modeling}, the absolute resistivity remains of the order of $\approx$ 1 $ \mu \Omega$ cm, and its low temperature value remains comparable to that at high temperature. Therefore the behavior does not represent a conventional metal–insulator transition or thermally activated insulating state \cite{mott2012electronic}.
In comparison, pure Cr${_\mathrm{2}}$Se${_\mathrm{3}}$ maintains complete metallic behavior throughout this entire temperature range \cite{wu2020magnetotransport}. 
The resistivity anomaly suggests the emergence of an additional scattering- or mobility-dependent contribution, possibly arising from short-range magnetic correlations, iodine-induced chemical disorder, redistribution among electronic channels with different mobilities, or subtle lattice distortions. In addition, as the material is cooled further below $T_N$, the resistance levels out (Figure \ref{BMP}(a)).

The weak nonlinear transport—the third-harmonic voltage signal, which reflects the intrinsic electronic inhomogeneities in a system \cite{das2012magnetically}, is measured on the same sample. This measurement is performed using a standard lock-in technique at a frequency of 37 Hz in normal four probe electrical transport geometry. The inset of Figure \ref{BMP}(a) displays the temperature dependence of the third-harmonic resistivity. A similar anomaly in the third-harmonic signal is consistent with the development of a spatially nonuniform microscopic current distribution near the crossover around 135 K. Such inhomogeneity could arise from spatial variations in iodine concentration, native defects, or locally varying electronic and magnetic scattering. This interpretation is also compatible with the nonlinear Hall response observed at 100 K, which indicates that a single dominant carrier-channel description becomes inadequate below the crossover and suggests the involvement of multiple transport channels with different mobilities. However, the present measurement does not uniquely identify the origin of the nonlinear response, since contributions from Joule heating and contact-related nonlinearities cannot be excluded. Comprehensive field-dependent nonlinear transport or fluctuation spectroscopy studies are still necessary, especially in the context of random resistor networks \cite{dubson1989measurement}. The characteristic anomalies in the dc resistivity and third-harmonic response occur at slightly different temperatures as mentioned. This difference may partly arise from the broad nature of both features and the associated uncertainty in determining their characteristic temperatures. In addition, the dc resistivity minimum may mark the onset of the transport crossover, whereas the third-harmonic anomaly may correspond to the temperature at which spatially inhomogeneous or nonlinear transport becomes most pronounced. Possible contributions from thermal lag, Joule heating, and contact-related nonlinearities should also be considered.

Figure \ref{BMP}(b) presents the field-dependent heat capacity, $C_p$, as a function of temperature. At 0 kOe, a distinct peak is observed near $T_N$, confirming the long-range magnetic ordering transition. In addition, a weaker change in the slope of the $C_p(T)$ curve is evident near $T^{*}$ (at 156 K). These characteristic temperatures are in good agreement with the temperature range established by other measurements. Upon applying a magnetic field of 30 kOe, the magnetic ordering temperature ($T_N$) remains essentially unchanged. However, a weak suppression of $C_p$ is observed at higher temperatures in the vicinity of $T^{*}$. This reduction in heat capacity may indicate a field-induced suppression of magnetic fluctuations and a corresponding decrease in magnetic entropy.

\subsection{Raman spectroscopy}

To examine whether the crossover near T* is accompanied by changes in the crystal lattice, temperature dependent Raman spectroscopy was performed over the temperature range 100-340 K. Figure \ref{BMP}(c) shows the Raman spectrum of I@Cr$_{\mathrm{2}}$Se$_{\mathrm{3}}$ sample at 300 K. Total six Raman-active modes are appeared near 120.6, 156.7, 173.6, 190.2, 204.5, and 245.9 $cm^{-1}$, labelled as $P_{1}$ to $P_{6}$ in the Figure. The recorded Raman spectra at different temperatures were fitted using Lorentzian line shapes to extract the frequency and linewidth of individual Raman modes. The temperature dependence of the phonon modes was also analyzed using a conventional three-phonon anharmonic model \cite{ghosh2021spin}, with the corresponding fits and the extracted peak positions as functions of temperature are provided in the Supporting Information (Figure S3). Figure \ref{BMP}(d) presents a temperature–Raman-shift contour map, showing a continuous evolution of all observed Raman features over the investigated temperature range.
Thus the Raman measurements provide no evidence for a symmetry-changing structural phase transition \citep{bera2023raman} or strong anomalous phonon renormalization \cite{ghosh2021spin} associated with the transport crossover. However, the absence of a Raman-resolvable anomaly does not exclude weak or mode-selective spin–phonon interactions below the experimental sensitivity.

\subsection{Theoretical results and Angle-resolved photoemission spectroscopy (ARPES)}

\begin{figure*}[ht]
\centering
\includegraphics[width=1\linewidth]{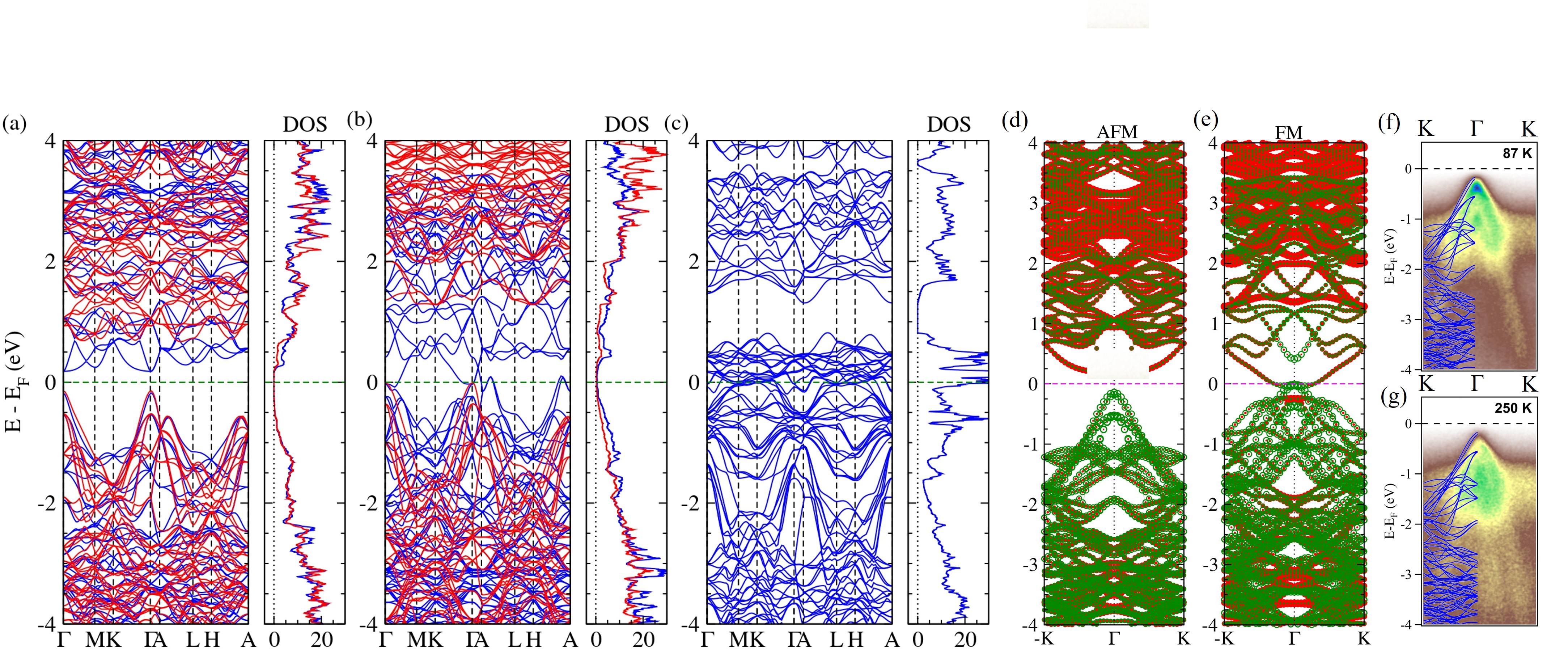}
\caption{Calculated band structure and density of states (in units of states/eV/spin) for (a) antiferromagnetic (AFM), (b) ferromagnetic (FM), and (c) nonmagnetic (NM) spin configurations. The red and blue lines represent the spin-up and spin-down components of the band structure and density of states, respectively. The green dashed line indicates the Fermi level. (d), (e) Calculated orbital contributions of Cr 3d states (green dotted lines) and Se 4p states (red dotted lines) to the band structure of pristine Cr${_\mathrm{2}}$Se${_\mathrm{3}}$ along the K-$\Gamma$-K direction for AFM and FM spin configurations, respectively.  (f), (g) ARPES spectra of in situ cleaved iodine-incorporated Cr${_\mathrm{2}}$Se${_\mathrm{3}}$ at 87 K and 250 K, respectively, showing the band dispersions along the K-$\Gamma$-K direction. The theoretical band structure for the AFM configuration is overlaid for comparison.}
\label{fig-DFTbands}
\end{figure*}

Figure \ref{fig-DFTbands} (a)-(c) depict the first-principles electronic structures (see Computational Details) of pristine Cr${_\mathrm{2}}$Se${_\mathrm{3}}$ for AFM, FM, and nonmagnetic (NM) spin configurations. In these calculations, the magnetic moments of Cr are aligned out-of-plane, consistent with experimental observations. Note that the magnetic anisotropy orientates from in-plane to out-of-plane as the lattice constant increases from isostructural Cr${_\mathrm{2}}$Se${_\mathrm{3}}$ to Cr${_\mathrm{2}}$Te${_\mathrm{3}}$ crystals~\cite{gebredingle2022first}. In our study, iodine incorporation induces lattice expansion and electronic doping, leading to out-of-plane magnetic anisotropy in Cr${_\mathrm{2}}$Se${_\mathrm{3}}$. Compensated AFM (cAFM) spin configuration both in-layer and intra-layer~\cite{gebredingle2022first} are found be lowest in energy in comparison to the other possible AFM configurations. To avoid iodine clustering in the unit cell and reduce computational cost in the supercell, the rigid band approximation is used to appropriately shift the Fermi level, thereby accounting for the effect of 8.33\% iodine doping by adjusting the total number of electrons in the doped system, as indicated by green dashed line in the Figure \ref{fig-DFTbands} (a)-(c). Our DFT calculations reveal that the cAFM, FM, and PM phases of Cr${_\mathrm{2}}$Se${_\mathrm{3}}$ are semiconducting, half-metallic, and metallic, respectively. These results align with the experimentally observed variation in resistivity across a broad temperature range. In particular, the low density of states near the Fermi level and the Dirac nodal line-like topological feature along the $\Gamma-A$ $k$-path in the electronic structure for the FM spin-configuration may be the possible origin of the observed negative magnetoresistance (MR). This negative MR has become a hallmark of several topological metals, such as WTe$_2$~\cite{wang2016gate} and Cd$_3$As$_2$~\cite{li2016negative}, TbPdBi~\cite{zhang2023robust}, etc.

Figures \ref{fig-DFTbands}(d) and (e) show the calculated orbital-resolved band structures of pristine Cr${_\mathrm{2}}$Se${_\mathrm{3}}$ along the K-$\Gamma$-K direction in AFM and FM spin configurations, respectively. The Cr 3d states (green dotted lines) dominate near the Fermi level, and the Se 4p states (red dotted lines) lie well below $E_F$. Previous first-principles calculations of Cr${_\mathrm{2}}$Se${_\mathrm{3}}$ \cite{gebredingle2022first} have shown that spin-down d-states near $E_F$ originate primarily from Cr(I) sites, while spin-up states originate from the Cr(II) and Cr(III) sites, with negligible Se 4p orbital involvement, which is consistent with our first principles calculations.  

Figures \ref{fig-DFTbands}(f) and (g) show ARPES spectra of in situ cleaved high-quality I-incorporated Cr${_\mathrm{2}}$Se${_\mathrm{3}}$ single crystals taken along the K-$\Gamma$-K direction. The spectra were measured at 87 K and 250 K with a photon energy of 445 eV. At 87 K, the spectra reveal a well-defined dispersive band centered at the $\Gamma$ point just below $E_F$, together with a weakly dispersing, nearly flat band spanning -1.8 to -0.8 eV binding energy. This is indicative of localized states with a large effective mass. At 250 K, the overall band dispersion remains intact; the most prominent effect is thermal broadening of spectral features, with no detectable redistribution of states near $E_F$.
The experimental dispersions were compared with the calculated electronic structures. The AFM calculation reproduces the overall shape of the dominant dispersive feature near $E_F$ more closely than the FM calculation, although several experimental features are not fully captured. The comparison is therefore qualitatively consistent with the predominantly antiferromagnetic interactions inferred from the magnetic measurements. Furthermore, the calculations describe the pristine host within a rigid-band approximation and do not explicitly include iodine-related disorder, local structural relaxation, or spatial variations in the electronic potential. The positioning of the Se 4p states far below $E_F$ leads to minimal hybridization with the Cr d-states. 
At both temperatures, the absence of a clearly resolved band crossing at $E_F$ is consistent with a semiconducting-like electronic structure. Nevertheless, small Fermi-surface pockets, weakly dispersing bands, or conducting states with low photoemission spectral weight may remain unresolved. Such states could contribute disproportionately to bulk charge transport and may help account for the unusually high electrical conductivity. Importantly, independent ARPES measurements performed at the DESY and Elettra beamlines on two different crystals revealed the same qualitative electronic structure, demonstrating the reproducibility of the observed behavior.

\section{Discussion}
Iodine incorporation into Cr$_2$Se$_3$ modifies the coupled magnetic and electronic response without changing in its average crystallographic symmetry. The magnetic susceptibility, heat capacity, and ESR measurements consistently identify an AFM-dominated ordered state below $T_N$ $\approx$ 52 K. A second, broader temperature anomaly appears near T* $\approx$ 150 K, where the susceptibility deviates from high-temperature Curie-Weiss behavior and the zero-field resistivity develops a shallow minimum. Unlike the sharp heat-capacity anomaly at $T_N$, the response near T* is broad and weakly field dependent, indicating a crossover involving short-range magnetic correlations or magnetic inhomogeneity rather than a second well-defined long-range ordering transition.

Although $\frac{d\rho}{dT}$ changes sign below the resistivity minimum, the absolute resistivity remains of the order of 1 $\mu\Omega\cdot\mathrm{cm}$, excluding a conventional metal–insulator transition or thermally activated insulating regime. The crossover is therefore more naturally attributed to an additional scattering- or mobility-dependent contribution that develops on cooling. The evolution from an approximately linear Hall response at 150 and 200 K to nonlinear Hall behavior at 100 K indicates that a single dominant carrier-channel description becomes inadequate below T*. In addition, the accompanying third-harmonic anomaly is compatible with an increasingly nonuniform microscopic current distribution. Collectively, these results suggest that iodine-related chemical disorder, native defects, magnetic correlations, and the redistribution of conductivity among channels with different mobilities contribute to the broad transport crossover. The absence of abrupt Raman or ARPES changes further indicates that this crossover is not driven by a symmetry-changing structural transition or a large reconstruction of the occupied electronic bands.

In the correlated paramagnetic regime, the approximately quadratic negative MR is consistent with field suppression of spin-dependent scattering. Within the magnetically ordered state, however, the much larger MR and its nonmonotonic temperature dependence point to a separate field-sensitive process involving the reorganization of the magnetic state and suppression of spin- or domain-dependent scattering. Therefore, the low-temperature giant MR need not arise from the same microscopic process responsible for the resistivity minimum near T*, although both demonstrate strong coupling between the magnetic and electronic degrees of freedom.

Several experimental observations are consistent with the emergence of short-range magnetic correlations over an extended temperature range above the long-range ordering temperature. This interpretation is supported by the nonlinear magnetic response above $T_N$, the weak ESR feature, and the broad field-dependent contribution to the heat capacity. These correlations may arise from competing exchange interactions, iodine-related chemical disorder, native defects, or spatial variations in the local magnetic environment.

A magnetic-polaron scenario provides one possible microscopic connection between these short-range magnetic correlations and the accompanying transport crossover \cite{kasuya1970magnetic,emin2013polarons,holstein2000polarons}. In this picture, mobile carriers become exchange-coupled to nearby Cr moments and locally modify or enhance their magnetic correlations, producing carrier-centered regions whose mobility and magnetic response differ from those of the surrounding. The formation or growth of such regions could contribute to the additional resistive component below T*, the nonlinear Hall response, the enhanced third-harmonic electrical signal, and the negative magnetoresistance observed above $T_N$. An applied magnetic field may partially polarize or reorganize these correlated regions, thereby reducing spin-dependent scattering and increasing carrier mobility. Local magnetic probes, such as $\mu$SR and neutron scattering, together with systematic field-dependent nonlinear transport measurements, will be required to test this scenario directly and to determine the characteristic size, binding, dynamics, and spatial distribution of any polaronic states in this material.

\section{Acknowledgments}
BD would like to thank Prof. Achintya Singha, Dr. Koushik Dey, Dr. Sumit Kumar Dutta, Dr. Sourav Bera, Mr. Arghyadip Bhowmik, and Dr. Satyabrata Bera for the experimental help and data analysis. BD is grateful to IACS for the fellowship. SD acknowledges the financial support from DST-SERB grant Nos. SCP/2022/000411 and CRG/2021/004334. SD also acknowledges support from the Central Scientific Service (CSS) and the Technical Research Centre (TRC), IACS, Kolkata. The authors thank the beamline staff Dr. Meng-Jie Huang and Dr. Jens Buck at the ASPHERE-III end-station of beamline P04, PETRA III, DESY, for their valuable technical assistance during the experiments. We thank DESY (Hamburg, Germany), a member of the Helmholtz Association (HGF), for access to their experimental facilities. 
We also acknowledge the financial support provided by the Department of Science and Technology (DST), Government of India, under the India@DESY collaboration initiative.
 
\section{Author contributions}
BD and SD conceived the project. BD carried out the single crystal growth, basic characterization, magnetization and transport measurements. KD conducted the theoretical calculations. OL, NGJA, and PN performed the ESR measurements and analyzed the ESR data. SM assisted with the Raman spectroscopy measurements. BD also received support from SMandal during the sample characterization. SM assisted with the Raman measurements. SC, TK, AB, KR and SKM conducted the ARPES measurements, with SKM analyzing the ARPES data. SN contributed to the specific heat measurement. RM assisted with the magnetotransport measurements. BD and SMandal wrote the manuscript taking input from KD, OL, SKM, SD and others.

\section{References}
\bibliography{references}

\end {document}